# Automated high-resolution backscattered-electron imaging at macroscopic scale

Zhiyuan Lang[1], Zunshuai Zhang[1], Lei Wang[1], Yuhan Liu[1,2], Weixiong Qian[1], Shenghua Zhou[1], Ying Jiang[3], Tongyi Zhang[1], and Jiong Yang[1]

[1]*Materials Genome Institute, Shanghai Engineering Research Center for Integrated Circuits and Advanced Display Materials, Shanghai University, Shanghai 200444, China*

[2]*Qianweichang College, Shanghai University, Shanghai, 200444, China*

[3]*School of Materials Science and Engineering, Zhejiang University, Hangzhou, 310023, China*

Email: yjiang0209@zju.edu.cn; jiongy@t.shu.edu.cn

## Abstract

Scanning electron microscopy (SEM) has been widely utilized in the field of materials science due to its significant advantages, such as large depth of field, wide field of view, and excellent stereoscopic imaging. However, at high magnification, the limited imaging range in SEM cannot cover all the possible inhomogeneous microstructures. In this research, we propose a novel approach for generating high-resolution SEM images across multiple scales, enabling a single image to capture physical dimensions at the centimeter level while preserving submicron-level details. We adopted the SEM imaging on the $AlCoCrFeNi_{2.1}$ eutectic high entropy alloy (EHEA) as an example. SEM videos and image stitching are combined to fulfill this goal, and the video-extracted low-definition (LD) images are clarified by a well-trained denoising model. Furthermore, we segment the macroscopic image of the EHEA, and area of various microstructures are distinguished. Combining the segmentation results and hardness experiments, we found that the hardness is positively correlated with the content of body-centered cubic (BCC) phase, negatively correlated with the lamella width, and the relationship with the proportion of lamellar structures was not significant. Our work provides a feasible solution to generate macroscopic images based on SEMs for further analysis of the correlations between the microstructures and spatial distribution, and can be widely applied to other types of microscope.

# Introduction

Microstructure of a material referring to its local composition, grains, phases, and other structural features at a microscopic scale plays a crucial role in determining the material's properties, such as mechanical strength, electrical conductivity, thermal conductivity, corrosion resistance, and more. One classical example is the material strengthening through manipulation of structural heterogeneity at various scales exemplified by impurity atoms, dislocations, twinning, grain boundary), precipitation/dispersion phases, composites, etc. Therefore, researchers have devoted considerable effort over an extended period to the characterization and design of microstructures that are intricately tied to the manufacturing process. Significant progresses have been made in the fabrication of metals with various microstructures using different compositions, manufacturing methods and parameters, and optimum microstructures could be determined by comparing the corresponding properties[1–3]. On the other hand, gradient materials were intentionally designed to establish the local microstructure-property correlation to screen the microstructure that exhibits the desired properties[4,5]. Regardless of the approach taken, it is crucial to accurately determine the relationship between the local microstructural characteristics and its properties throughout the entire sample. Even in the first approach, microstructural gradient or structural non-uniformity can exist within a single ingot due to local variations in parameters during fabrications[6,7]. This phenomenon of structural non-uniformity and anisotropy is more common in materials prepared by additive manufacturing, which builds parts by adding material one layer at a time and involves complex cyclic thermal history[8,9]. Therefore, characterizing the structure of micro-regions to understand the performance of the entire sample may be inaccurate and far from sufficient. High-throughput characterization method to reveal the microstructures over large length scales is indispensable to illustrate the structural heterogeneity within the ingot, understand the local microstructure-property correlation, and screen the superior microstructure in a rapid way.

We take the common material characterization technique, scanning electron microscopy (SEM), as an example. SEM plays an extremely significant role in the characterization of microstructures, providing detailed imaging and analysis of surface morphology, grain boundaries, phase distribution with high-definition (HD) and magnification. However, obtaining cross-scale images through SEM is often an impossible task due to limitations in resolution and working distance. Higher resolution imaging often requires shorter working distances, which restrict the field of view and the ability to image large areas. Therefore, it poses a challenge in simultaneously optimizing imaging resolution, field of view, and imaging speed. One possible solution is super-resolution methods[10–15] based on deep learning, which have shown tremendous potential in enhancing image resolution. These methods utilize deep neural networks to learn the mapping between low-resolution (LR) and high-resolution (HR) images, thereby generating HR images from LR inputs. Dong et al.'s[16] SRCNN is considered the seminal work in image super-resolution reconstruction based on deep learning. Kim et al.'s[17] VDSR introduced residual learning. Ledig et al.[18] utilized generative adversarial networks for super-resolution. Zhang et al.[19] enhanced feature learning with channel attention and proposed the residual in residual structure. Liang et al.[20] utilized Swin Transformer for image super-resolution, combining Transformer with CNN. These methods has been successfully applied in various fields, including medical imaging[21–24] and remote sensing[25–27]. In the field of electron microscopy, Orkun Fura et al.[28] employed generative adversarial networks to enhance the resolution of SEM images of fractured cathode materials. Devendra K. Jangid et al.[29] found that incorporating domain knowledge into the training process of super-resolution models improves their performance on electron microscopy image datasets. However, existing work has limitations in resolution enhancement while maintaining the original field of view, typically achieving 2x, 4x, 8x, or 16x enhancements. As magnification increases, the "authenticity" of the images diminishes. Even with a 16x magnification, it falls short of the requirements for cross-scale imaging. Moreover, limited magnification may not meet the requirements when characterizing materials at the centimeter scale while preserving microscale details, which cannot be solely addressed by conventional super-resolution techniques. Another potential solution for cross-scale imaging is image stitching. Wenjing Yin et al.[30] have developed an all-weather continuous autonomous imaging system for transmission electron microscopy, enabling high-throughput image acquisition at the petabyte scale through parallelization and automation. This work achieved high-throughput imaging and stitching to obtain extremely large images. However, it suffers from fatal drawbacks such as equipment modifications and excessively long acquisition times.

To overcome these challenges, we propose a novel method for generating cross-scale HR SEM images, i.e., macroscopic ones with submicron-level details. The required equipment is simply a standard electron microscope with video recording capability, along with our plug-and-play system called the Cross-Scale Electron Microscopy Image Generation System (CEMI), as shown in Figure 1. Given a LD SEM video as input, CEMI extracts consecutive LD frames, and feeds them into a pre-trained denoising model to generate corresponding HD images. The images are then stitched together using an image stitching module. This technique holds paramount significance for materials research. Firstly, our proposed method addresses the limited imaging range issue of traditional SEM techniques, allowing researchers to obtain images over a broader range. By generating large-scale HR images, researchers can better observe the microstructure and properties of materials, providing a more comprehensive understanding in the field of materials science. Secondly, we greatly reduce the cost of acquiring HD SEM images based on LD ones at much faster scanning speed. Furthermore, we explore analytical methods for the cross-scale image to gain better insights into the microstructure and properties of materials. Due to the enormous imaging range of cross-scale images, it is challenging for the human eye to derive meaningful conclusions directly. Therefore, we employed image segmentation techniques to segment the microstructures of interest. In this work, we conducted a statistical analysis of the distribution of three microstructural components in $AlCoCrFeNi_{2.1}$ EHEA. These components include lamellar structures, lamellar width, and the body-centered cubic (BCC) phase. Furthermore, we analyzed their correlation with material hardness. CEMI is not only applicable to SEM but also easily integrable into various types of microscopes, making it a valuable tool for researchers across different scientific disciplines.

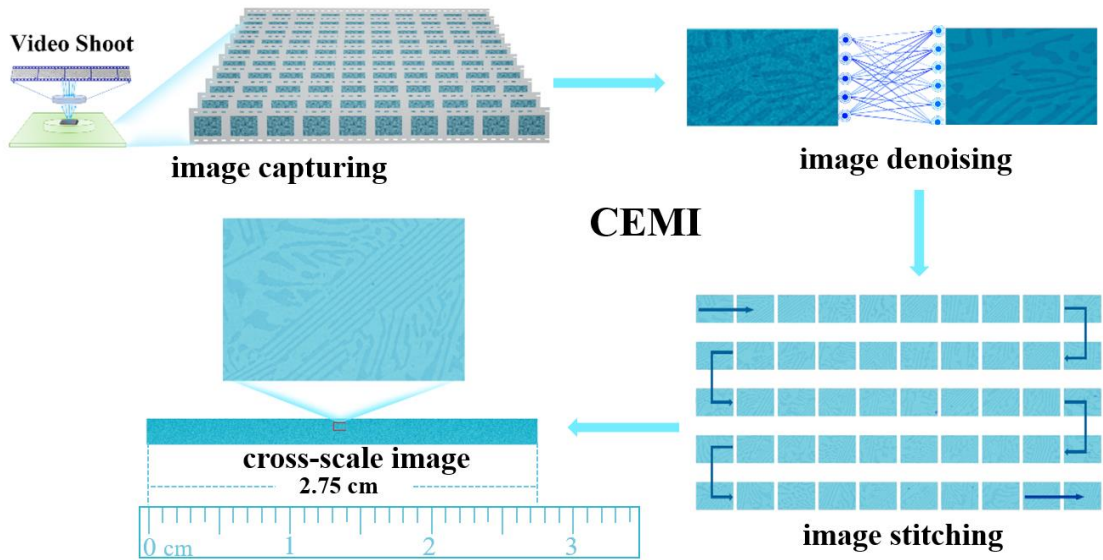

**Figure 1. Overall workflow of Cross-Scale Electron Microscopy Image Generation System and its modules.**

## Results

**Image capturing.** Conventionally, a series of high-magnification photographs with partial (typically 30-50%) overlaps must be captured to serve as input data for image stitching over large length scales[31–36]. This is also true for the commercial software Thermo Scientific Maps, which is only compatible with its own facilities. However, shooting HD images are very time-consuming and labor-intensive. Even for skilled operators, taking a HD SEM image requires 1-2 minutes (including locating the local region, focusing, scanning and saving). By contrast, CEMI offers flexibility to deal with either low-quality videos or LD images directly. Compared with HD image (cycle time=26.2 s), it only takes one tenth of the corresponding time to acquire the LD image (cycle time=2.7 s). Furthermore, we can continue to save time in data acquisition and free up manpower by using the automatic sample stage translation and video capture function. By setting an appropriate movement velocity, a low-quality video containing structural information throughout the lateral x-axis could be obtained. In this way, we finally collected 18 videos, from which 3902 frames were automatedly extracted by setting appropriate extraction intervals, covering the specimen with a macro size of 2.75 cm x 0.175 cm as shown in Figure 2a. To facilitate subsequent stitching, the extracted adjacent video frames have a certain degree of overlap (which can be cumbersome to manually control). In fact, we can continue to magnify the image to capture videos, obtaining more photographs and finer structural information even on the same order of the SEM definition. After preprocessing by the image acquisition module, the LD image is sent to the denoising module for denoising to obtain the corresponding HD image.

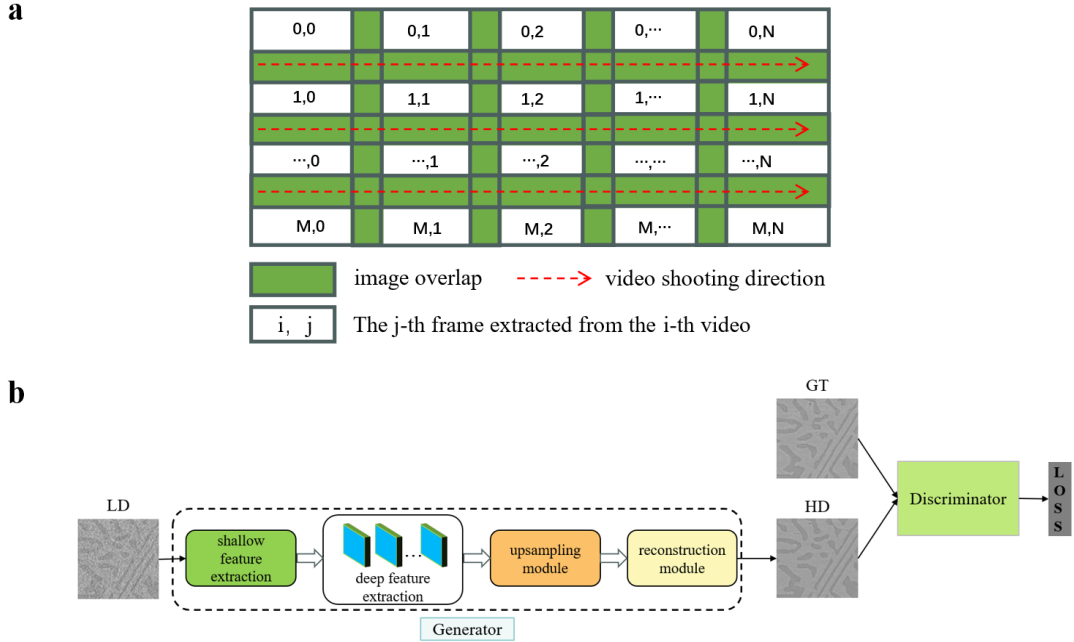

**Figure 2. a** Illustrates the SEM imaging path and the frame extraction process for video-based image acquisition. **b** Schematic diagram of the denoising model's structure

**Image denoising.** The overall structure of the denoising model is inspired by ESRGAN[37], which can be used for image denoising when the magnification factor is 1, and utilizes an adversarial neural network[38] consisting of a generator and a discriminator. As shown in Figure 2b, the generator consists of four components: shallow feature extraction, deep feature extraction, upsampling module, and reconstruction module. The deep feature extraction consists of multiple basic blocks, which are implemented based on the Residual-in-Residual Dense Block[37]. Each basic block contains three dense blocks, and each dense block consists of five convolutional layers (with ReLU activation applied after the first four layers). The discriminator employs a U-Net[39] network with spectral normalization regularization[40], which helps stabilize the training process. We trained the model using 100 pairs of high-low definition images of high-entropy alloy samples (detailed shooting methods can be found in the Methods section). Due to the instability of adversarial training, we first trained the generator separately using L1 loss. The resulting denoising model is referred to as SEMNET. Then, we used the trained SEMNET as the initialization for the generator in the adversarial neural network and combined L1 loss, content perception loss, and adversarial loss to obtain the final denoising model, named SEMGAN. The results are shown in Figure 3a. Overall, images generated by SEMNET tend to be relatively smooth, with some loss of fine details compared to HD real images. In contrast, SEMGAN performs better in this regard and is therefore used as the denoising model in CEMI. The so-obtained HD images show better similarity with the ground-truth (GT) HD images, than the LD images. For example, we evaluated the numerical disparities in BCC phase fractions (details for this microstructure will be discussed later in this work) between the output images from the denoising model and the real 100 HD-LD image pairs in the training set (Figure S1). The results of the denoising model's output images closely align with those of real HD images, while real LD images show larger disparities from the other two. Furthermore, when only HD images are available, we can use a degradation model to obtain the corresponding LD images, which can then be used for training the denoising model. You can find more detailed information in the Discussion section.

The denoising module is an indispensable component of CEMI in this EHEA case for two main reasons. Firstly, for the purpose of rapidly and cost-effectively generating cross-scale images, low-definition images obtained through the image capture module must undergo the denoising module before being used for subsequent stitching. Secondly, the denoising module can be used independently, reducing the cost of acquiring high-definition images. Furthermore, as per our knowledge, there is currently no high and low-definition SEM image data set for training denoising models. We have produced and made the above data public, which will help promote the research of denoising models dedicated to SEM area.

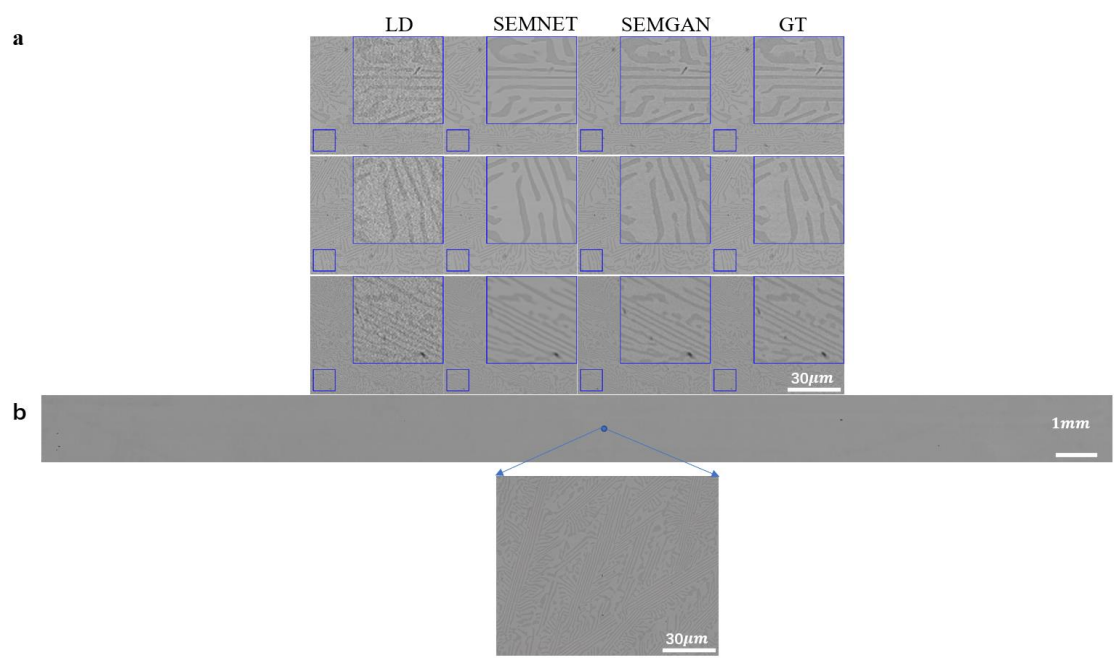

**Figure 3. Denoising model performance demonstration. a** LD represents the low-definition input image to the model, followed by the HD images generated by SEMNET and SEMGAN. GT denotes the GT HD image. **b** The image shows a large-scale image created by stitching together 3902 images. It has a resolution of 123672 x 7848 pixels and a physical size of 2.75 cm x 0.175 cm. The actual length of the material is approximately 28 millimeters.

**Image stitching.** By using the trained denoising model, we can input LD images obtained from the image acquisition module and generate corresponding HD images. Subsequently, we employ image stitching methods to merge the generated HD images and create cross-scale images. The stitching process takes time proportional to the number of images being stitched, with approximately 3 hours needed for the 3902 images. In our experiments, we found that LD images are not suitable for large-scale image stitching due to their limited feature points. Even when combining two LD images, there is still a possibility of stitching failure. Therefore, LD images are not usable when the number of images for stitching reaches thousands. As shown in Figure 3b, we generated a large-sized image of 123672 x 7848 pixels using the 3902 images, which corresponds to a physical size of 2.75 cm x 0.175 cm. An animation of the zooming-in process on the cross-scale image is demonstrated in Gif S1(A.gif). For the ease of description in the following text, we will refer to the horizontal direction as the x-direction and the vertical direction as the y-direction. Remarkably, even at this scale, we were able to observe details at the submicron level. The amplification level in this work is equivalent to reading a textbook with font size 10 from a distance of 300 meters.

The significance of cross-scale images lies in their ability to capture a wide range of information that is not easily obtained through conventional imaging techniques. In our experiments, we utilized SEM, but any microscope capable of recording videos can leverage CEMI to generate cross-scale images that were previously unattainable. From a materials application perspective, overwhelming amount of information provided by cross-scale images makes manual analysis practically impossible. Consequently, there is a need to develop CEMI plugins for automated cross-scale image analysis. In the following section, we demonstrate the scientific value of CEMI by showcasing a plugin for image segmentation in high-entropy alloys.

**Applications of Cross-Scale Image.** In the context of cross-scale image generated through stitching, a more in-depth exploration of analytical methods has been undertaken. The sample we adopted here was $AlCoCrFeNi_{2.1}$ EHEA, in which fine, intricately spaced phases provide exceptional mechanical properties, making it gain significant attention in aerospace, automotive, and other applications in various industries. Prior investigations indicated the hardness, strength and ductility are correlated with some factors such as the contents of lamellar eutectic structures, BCC phases, and size of eutectic structures[41–45]. In order to illustrate the structural heterogeneity within the ingot and determine local microstructure-property correlation in a high throughput manner, we used CEMI to automatedly image the microstructure with a high resolution throughout the entire ingot. The ingot of $AlCoCrFeNi_{2.1}$ EHEA was manufactured by arc melting method, with a diameter of ~ 3 cm. To quantitatively obtain the distribution of different types of microstructural features, the cross-scale image has been partitioned into a grid of 19x243 smaller images, each measuring 508x413 pixels. Subsequently, dedicated procedures have been applied to these smaller images, encompassing lamellar structure segmentation, lamellar width quantification, and BCC phase proportion estimation. The outcomes of these operations are then visualized on a 19x243 matrix, where darker shades indicate higher numerical values in the respective regions. The so-obtained distribution of microstructures serves as a quantitative tool to eliminate the bias of inhomogeneity, and such information is not able to be obtained from single SEM (or other type) image.

For lamellar structure segmentation, an image segmentation model has been employed. In recent years, image segmentation networks[46–50] based on deep learning have experienced rapid development. In this work, the U-net[39] architecture, known for its robust segmentation capabilities, specifically a U-net++[51] variation, has been utilized for the lamellar segmentation task, as illustrated in Figure 4a. Based on the results of lamellar structure segmentation, the widths of the lamellar structures have been quantified, as demonstrated in Figure 4b. This was achieved by determining the minimum bounding rectangle for each lamellar region within the smaller images. The width of each lamellar region (represented by the length of the green line segment) and the count of lamellar structures (indicated by the number of black lines intersected by the green line segment) were computed. It's important to note that due to the presence of multiple lamellar regions within each smaller image, the width of lamellar structures was calculated separately for each region, and the average value was considered as the lamellar width within the respective smaller image. In the case of BCC phase proportion estimation, the original image was first transformed into a grayscale representation. Subsequently, a binary image was created through a thresholding procedure[52], enabling the quantification of the BCC phase's respective proportion (represented by the black regions in the binary image), as depicted in Figure 4c. The comprehensive distribution of these three microstructural characteristics is illustrated in Figure 4def. The proportion of lamellar structures is relatively low on the far right, with some degree of fluctuation in other parts. Lamellar width is higher at both ends and lower in the middle. The BCC phase exhibits a distinct feature of being higher in the middle and lower at both ends. The x and y axes in the distribution graph are the corresponding dimensions of the sample, allowing for a direct comparison with real measurements. This approach facilitates direct observation of the distribution of these microstructural features across different regions, thereby aiding subsequent analyses.

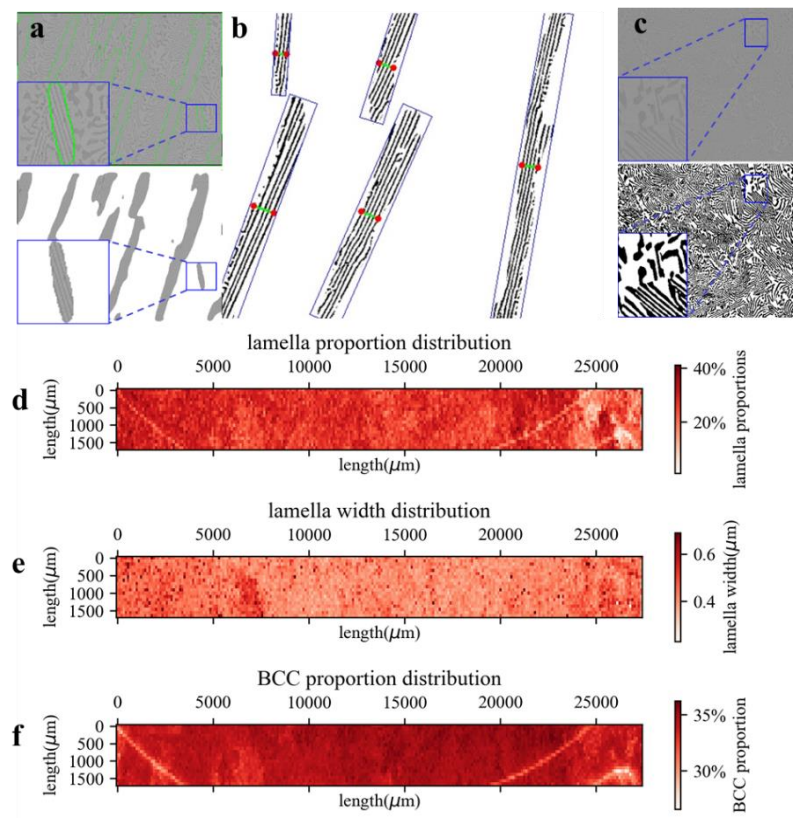

**Figure 4. Visualization and analysis of microstructures in cross-scale images. a.** The portion enclosed by green lines represents segmented lamellar components. **b.** Lamellar widths are calculated based on the lamellar segmentation results. **c.** Image binarization, black regions denote the BCC phase. **d.** Distribution of lamellar structures. **e.** Distribution of lamellar structure widths. **f.** Distribution of BCC phase composition.

Subsequently, we quantified the microstructural information at corresponding positions. This information was condensed into one dimension by averaging along the y-axis, as illustrated in Figure 5abc. The overall distribution of lamellar width ranges from 550 to 750 nanometers, predominantly concentrated around 600 nanometers. The overall distribution of the lamellar structure proportion ranges from 15% to 33%, with a main concentration near 26%. The distribution of the BCC phase ranges from 33% to 35%, predominantly centered around 34.5%, in accord with the phase fraction reported in the literature[53,54]. To evaluate the local mechanical properties varying with the structural features, nanoindentation experiments, a powerful method to investigate the surface mechanical properties, were carried out along the lateral x-axis to determine the hardness of small volume with small load and small tip size[55–57]. As depicted in Figure 5d, the hardness changes with the length along the x-axis, initially increasing from 450 HV to 550 HV and subsequently descending back to approximately 450 HV. Maximum harness is achieved in the range of 1.1-1.5 cm, that is, the middle section of the ingot with a smaller lamella width of 600 nanometers, lamella content of 26% and maximum BCC content of 35%. Based on the data from Figures 5abc, we calculated the Pearson correlation coefficients between hardness and each of the three microstructural characteristics, resulting in values of -0.4439, -0.2645, and 0.633, respectively. The feature of BCC, showing higher values in the middle and lower values at the ends, exhibits the strongest correlation with hardness, while the distribution of the lamellar structure demonstrates the weakest correlation. Even when using combinations of microstructural distributions, there was no significant improvement in the overall correlation with experimental hardness, and there was a tendency toward overfitting (see the Figure S2). We believe this is primarily due to fluctuations in the results of the indentation experiments and image segmentation, particularly in relation to the segmentation of lamellar structures. Based on the existing data, we conclude that the proportion of the BCC phase and the width of lamellar structures are the crucial factors influencing hardness.

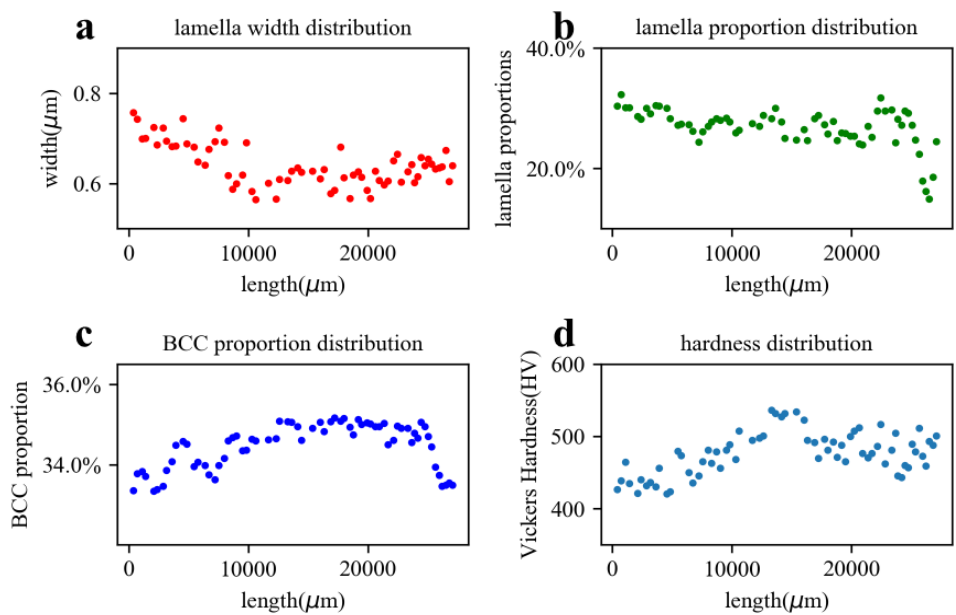

**Figure 5. Microstructure and hardness information.**

# Discussion

One of the key points of our work is the denoising model trained from 100 pairs of high-low definition images. However, taking pairs of high-low definition images has complicated post-processing operations such as image alignment, which will limit the generalization of CEMI. In addition, we found that in many electron microscopy laboratories, they have accumulated a lot of HD SEM images, but there are no corresponding LD images. To address the above situation, we explored the feasibility of using only HD images, and the LD images are generated by degradation model, and then used for denoising model training. The relationship between HD and LD images can be modeled using Equation (1), where x represents the LD image, y represents the existing HD image, k is the blur kernel, r is the downscaling factor, and n is the noise. The degradation model is complex and irreversible. Although classic degradation models can represent the degradation process, directly using them to generate LD images leads to limited diversity in the generated LD images. Inspired by Real-ESRGAN[13], we mix the degradation processes, as shown in Equation (2), randomly combining blur, down-sampling, and noise operations twice, with the possibility of skipping each step (For further details, please refer to the Methods section.). The model trained using the degradation data is referred as the "synthetic model", while the one trained with paired images is termed the "pairs model". As shown in Figure 6, the synthetic model effectively recovers primary textural details, but it may overlook smaller black point-like areas. The Learned Perceptual Image Patch Similarity (LPIPS)[58] values for the synthetic and pairs models vs the HD image in the testing dataset are 0.3363 and 0.2772, respectively. Lower LPIPS values indicate greater similarity between two images, while higher values indicate greater dissimilarity. While there is indeed some difference between the synthetic and pairs models, the synthetic model remains suitable for the denoising module in CEMI. We replicated the process illustrated in Figure 1 using the synthetic model and conducted a statistical analysis of the BCC phase. The results, as shown in Figure S3, although there are differences in values, the overall distribution trend is very similar to Figure 5c. The suitability of the synthetic model in CEMI is because subsequent applications and analyses primarily focus on the texture structure of the images rather than fine-grained pixel-level differences.

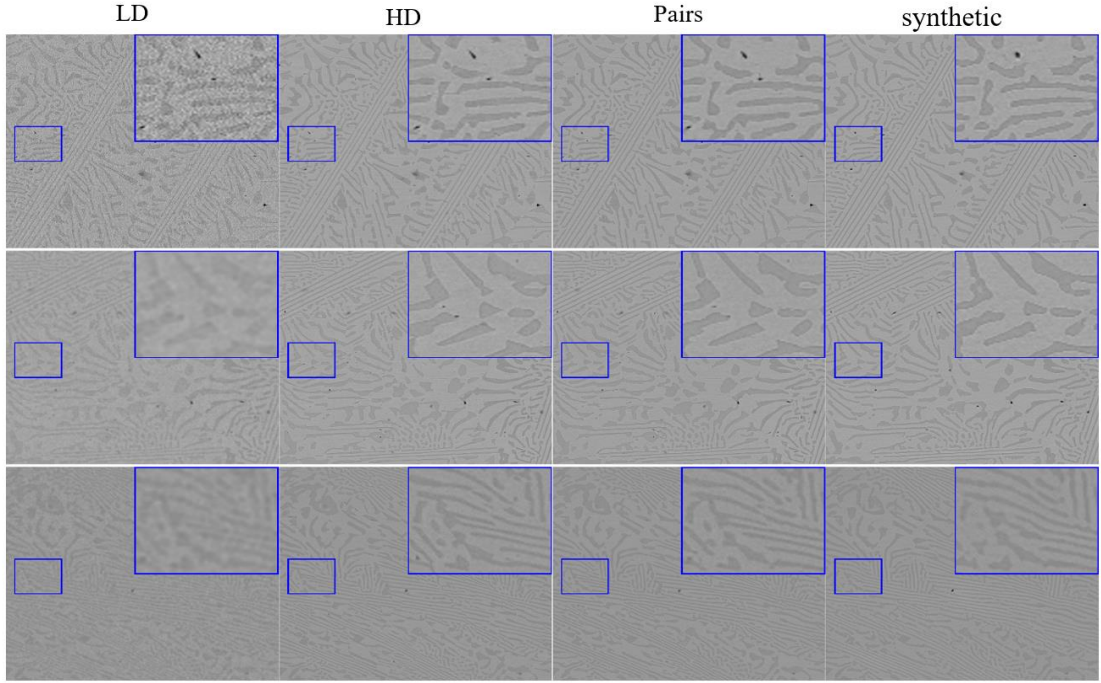

**Figure 6. Comparison of denoising results.** demonstrates the performance of SEMGAN trained on real paired data and degradation data. The "Pairs" section shows the results obtained with real paired data, while the "Synthetic" section presents the results obtained with degradation data.

$$x = D(y) = (y \otimes k)\downarrow_r + n \tag{1}$$

$$x = D^2(y) = (D_2 \circ D_1)(y) \tag{2}$$

In addition to the alloy materials studied in this work, CEMI can also be used to generate cross-scale images for other types of materials. We synthesized a series of gradient materials $(PbSe)_{1-x}(SrSe)_x$ with varying proportions of PbSe and SrSe at different locations. We captured a 6-minute video, extracted 63 HD images, and successfully stitched them into Figure S6, with a resolution size of 9096 x 608 pixels and a physical size of 2165.75μm x 145.3μm. The sample exhibits semiconducting properties, and this panoramic image can be used for further analysis of the component distribution at different positions. It is notable that due to the high clarity of the original video, no denoising module was required before stitching, demonstrating the flexibility of CEMI.

In this research, we have developed a cross-scale electron microscope images generation system based on computer vision techniques. Through the utilization of video frame extraction, denoising models (based on 100 pairs of LD and HD images), and image stitching, we have successfully generated electron microscope images ranging from centimeter to submicrometer scales. The overall function of CEMI is to generate cross-scale images, which can be applied when the horizontal axis has a meaningful physical scale, such as in the case of gradient structural materials. On the other hand, each module of CEMI can also be used individually to address the issue of material structural heterogeneity. For instance, one can directly capture a certain number of LD images, apply the denoising module in CEMI, and then perform image stitching to obtain the cross-scale image containing micrometer-level microstructures. We then explore methods for cross-scale image analysis using image segmentation model. Three different microstructures are examined, and the proportion of the BCC phase and the width of lamellar structures significantly impact hardness. The primary benefit of CEMI is its ability to rapidly and automatically generate cross-scale images with spatial distribution, which is of great significance in materials science research. Understanding the structure and properties of materials at different scales is crucial for developing new materials and improving existing ones. This technology aids in identifying and analyzing material defects and studying the effects of processing and treatment on material performance. Furthermore, its modular design ensures that CEMI can be continuously improved by incorporating the latest denoising, stitching, and segmentation techniques as deep learning technology advances.

**Methods**

**Materials preparing.** The AlCoCrFeNi$_{2.1}$ EHEA was prepared via arc melting high purity elements with purity > 99.99% in argon atmosphere[59,60]. Briefly, the metal blocks were mechanically ground to remove the surface oxide layer, cleaned by anhydrous ethanol solutions within an ultrasonic cleaning machine, and then dried with cold air before weighing. The alloy elements were sequentially placed in a crucible within the furnace based on their melting points. High purity argon gas was adopted for purging and protecting before and during manufacturing. The melting current was set to about 150 A.

**Sample polishing.** The AlCoCrFeNi$_{2.1}$ EHEA ingot was cut as shown by the schematic in Figure S4a. Prior to the microstructural characterizations, the ingot was respectively ground with 1000, 2000, 3000, 5000, and 7000 grit SiC papers. The polishing process continued until all unidirectional scratches on the surface became invisible. Subsequently, vibration polishing was performed using a VibroMet2 vibration polishing device from Buehler, USA, operating at 20% power for approximately 5 to 10 minutes. Finally, the AlCoCrFeNi$_{2.1}$ EHEA specimen was ultrasonicated in ethanol, and dried in air to acquire clean and fresh surfaces, as shown in Figure S4b.

**Image data sets and video shooting.** Microstructural heterogeneity of the AlCoCrFeNi$_{2.1}$ EHEA specimen was characterized within a field-emission scanning electron microscope (SEM, model G300, Carl Zeiss, Germany) at 500X magnification, with a 60 μm aperture size, contrast information set at 49.5%, and brightness at 68.8%. The backscattered electron detection (BSD) mode was utilized, and low, high-definition image data shots of the training set (containing 100 low-definition images and 100 high-definition images) were captured using Pixel Avg mode with a scan speed of 5 (Cycle time = 2.7 s) and Line Avg mode with a scan speed of 7 (Cycle time = 26.2 s), respectively, at the same site. Furthermore, a low-definition video was recorded along the X-axis at a moving speed of stage vector X = 2% in the Pixel Avg mode (Cycle time = 2.7 s) with all other conditions being the same. The sample table was returned to its initial position, moving a specific distance in the Y-axis direction, and then the movement continued along the X-axis at the same speed to capture the next video data. This process was repeated until panoramic video data of the macro sample was obtained.

**Nanoindentation experiment.** The microhardness of AlCoCrFeNi$_{2.1}$ EHEA was assessed using a nanoindentation instrument (model iMicro, KLA), with maximum indentation depth of 5000 nm, and maximum load of 1 N. To ensure accurate results, a minimum of 3 indentations were made at specific x-position, with sufficient distance between each indentation to avoid overlapping effects.

**Image processing.** For the dataset of 100 pairs of HD and LD images, we removed microscope parameters information unrelated to the images themselves. Then, we randomly selected 90 pairs for training the denoising model and 10 pairs for evaluation. From these pairs, we further selected 24 LD images and annotated the regions of interest corresponding to the lamellar components using Labelme[61] for training the segmentation model. Regarding the LD video data, we utilized OpenCV to extract frames with overlapping views, ensuring an overlap rate of 30% to 60% between adjacent frames.

**Denoising model.** We employed an adversarial neural network approach, where the generator's detailed structure is depicted in the provided diagram (Figure S5). The discriminator utilized a U-Net architecture. During training, the upsampling module had a magnification factor of 1, ensuring consistent input and output sizes. For the SEMNET training stage, we employed L1Loss, and for training SEMGAN, we initialized the parameters using the pre-trained SEMNET model. The loss function comprised adversarial loss, content perception loss, and L1 loss[13].

**Image degradation.** In our study, we employed Gaussian noise and Poisson noise with probabilities of 0.5 each to obtain LD images from the HD counterparts. The noise sigma range was set between 1 and 30, while the Poisson noise scale ranged from 0.05 to 3. For the second degradation process, the noise sigma range was adjusted to 1-25, and the Poisson noise scale was set between 0.05 and 2.5.

**Image stitching.** To merge the captured images seamlessly, we utilized PanoramaStudio 3.6.7 Pro for the image stitching process.

**Image segmentation model.** Our segmentation model was trained using a dataset of 21 HD images for training and 3 HD images for validation. We adopted the U-net++ architecture with the se_resnext50_32x4d encoder for the segmentation model.

**Training details.** We incorporated transfer learning in our denoising model training. Initially, we fine-tuned the SEMNET model based on the pre-trained ESRGAN[37] model to achieve faster convergence. Subsequently, we trained the SEMGAN model based on the improved SEMNET. During training, the batch size was set to 18, and we utilized Adam[62] optimizer with a learning rate of 1e-4. All models were trained for a total of 40,000 iterations. For implementation details not mentioned in the paper, we followed the guidelines provided by the ESRGAN[37], this module is developed based on the BasicSR[63] framework. The segmentation model employed a pre-trained model based on the imagenet[64] dataset and utilized Dice Loss[65] as the loss function. The optimizer used was Adam with a learning rate of 0.0001, and the training was conducted for 80 epochs. All training of deep neural networks was performed on a machine equipped with an NVIDIA RTX 3090 GPU.

**ACKNOWLEDGEMENTS**

This work is supported by the National Natural Science Foundation of China (Grant Nos. 52172216 and 92163212) and Shanghai Technical Service Center of Science and Engineering Computing, Shanghai University. J.Y. acknowledges the support from Hefei advanced computing center and Shanghai Engineering Research Center for Integrated Circuits and Advanced Display Materials. Y.J. acknowledges the support of Dr. P. J. Shi at Shanghai University for the alloy preparation.

# Supplementary Information

# for

# Automated high-resolution backscattered-electron imaging at macroscopic scale

Zhiyuan Lang[1], Zunshuai Zhang[1], Lei Wang[1], Yuhan Liu[1,2], Weixiong Qian[1], Shenghua Zhou[1], Ying Jiang[3], Tongyi Zhang[1], and Jiong Yang[1]

[1]*Materials Genome Institute, Shanghai Engineering Research Center for Integrated Circuits and Advanced Display Materials, Shanghai University, Shanghai 200444, China*

[2]*Qianweichang College, Shanghai University, Shanghai, 200444, China*

[3]*School of Materials Science and Engineering, Zhejiang University, Hangzhou, 310023, China*

Email: yjiang0209@zju.edu.cn; jiongy@t.shu.edu.cn

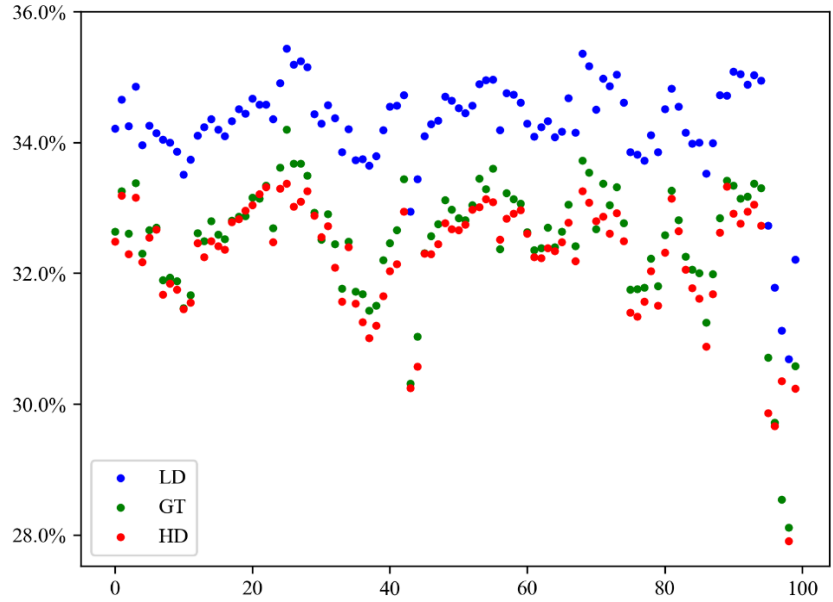

**Figure S1. The BCC phase statistics results for different image types.** LD represents low-definition images, GT denotes the ground-truth high-definition images, and HD corresponds to high-definition images generated using SEMGAN. The horizontal axis corresponds to the test image number, while the vertical axis indicates the corresponding phase fraction. We examined the differences in BCC phase fractions among different image types. Overall, the disparity between GT and HD is minimal, with errors averaging below 0.2%. LD exhibits larger differences from GT, averaging around 1.7%.

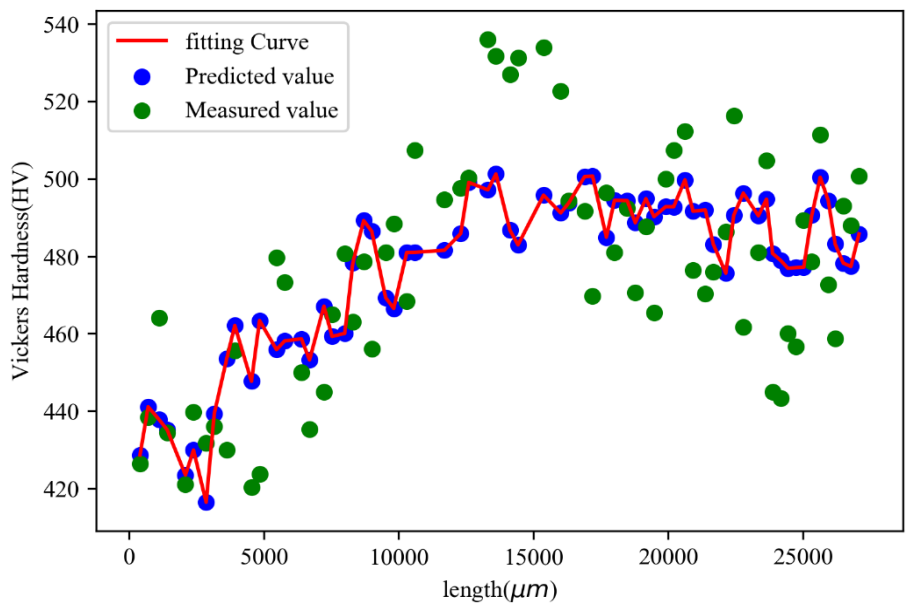

**Figure S2. Use lamellar structure distribution, BCC phase distribution and lamella width distribution to fit the hardness.** The correlation is 0.7154. Although it is improved compared to the BCC phase, there is an overfitting trend.

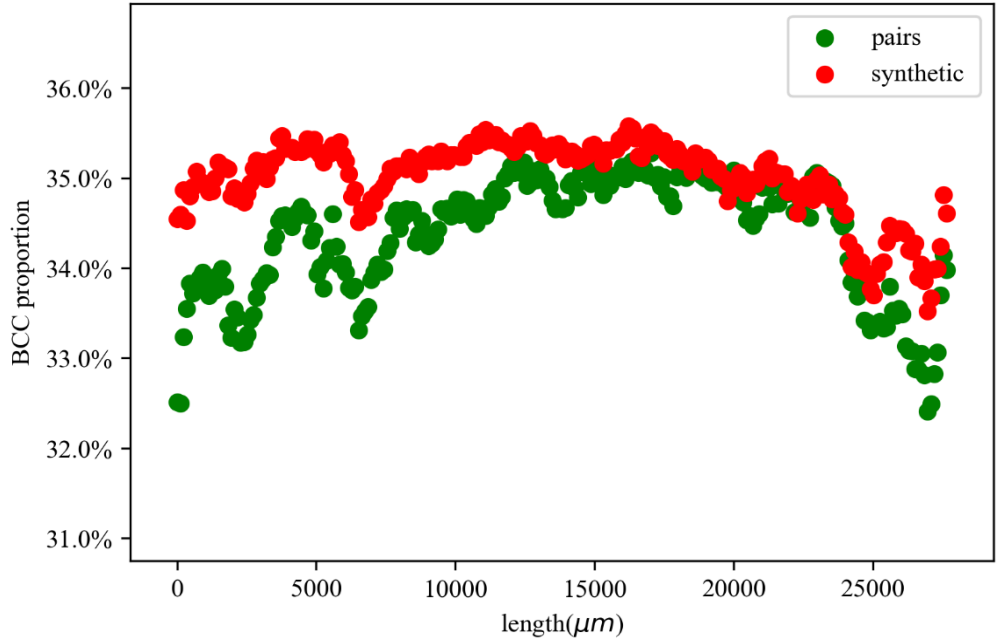

**Figure S3. BCC phase statistics are performed using pairs model and synthetic model respectively.** Although there is a numerical gap between synthetic and pairs, the overall distribution trend is basically maintained.

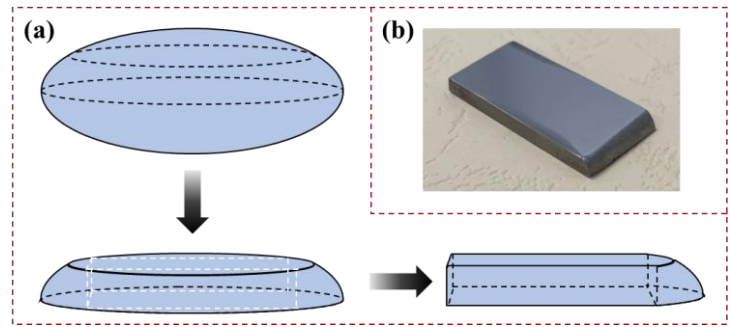

**Figure S4. a** Schematic diagram of the AlCoCrFeNi$_{2.1}$ EHEA sample. **b** Oblique view of the AlCoCrFeNi$_{2.1}$ EHEA ingot.

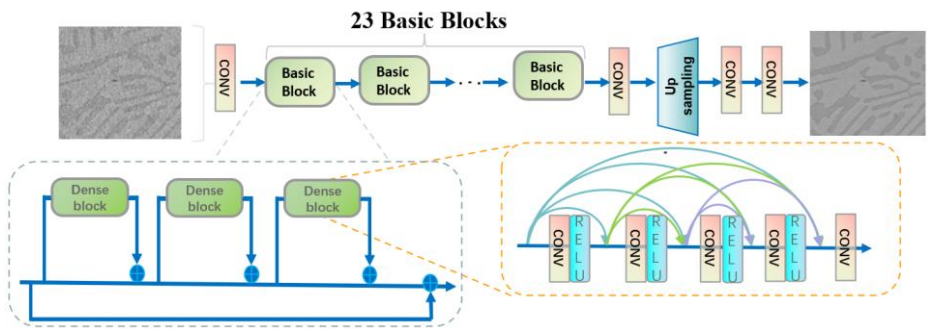

**Figure S5. Architecture of Generator Network**

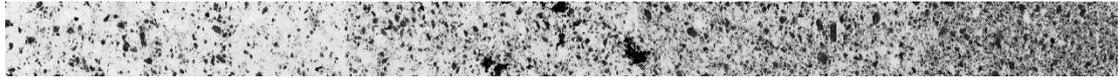

**Figure S6. The Electron Microscopy Image Stitching of PbSe-SrSe. It has a resolution of 9096 x 608 pixels and a physical size of 2165.75μm*145.3μm.**